\documentclass[article]{JHEP3}

\usepackage{amsmath,amssymb}
\usepackage[dvips]{graphics}
\usepackage{epsfig}

\newcommand{\be}{\begin{equation}}
\newcommand{\ee}{\end{equation}}
\newcommand{\bea}{\begin{eqnarray}}
\newcommand{\eea}{\end{eqnarray}}
\newcommand{\bean}{\begin{eqnarray*}}
\newcommand{\eean}{\end{eqnarray*}}

\relax

\def\d{\partial}

\def\a{\alpha'}

\def\R{\mathcal{R}}

\title{Scattering of spherically symmetric $d$--dimensional $\a$--corrected black holes in string theory}

\author{Filipe Moura
\\
Centro de Matem\'atica da Universidade do Minho, \\Escola de Ci\^encias, Campus de Gualtar, \\4710-057 Braga, Portugal\\
\\
\email{fmoura@math.uminho.pt}
}

\abstract{We study scattering of minimally coupled massless scalar fields by non--extremal spherically symmetric black holes in $d$ dimensions with string--theoretical $\a$ corrections. We then obtain a formula for the low frequency absorption cross section for every black hole of this kind, which we apply to known black hole solutions. We compare the $\a$ corrections for the absorption cross section with those for the black hole entropy, obtained through Wald's formula, in each case concluding that these corrections are different. We find a general covariant formula for the absorption cross section including $\a$ corrections, in terms of the horizon area and temperature.
}

\begin{document}



\vfill

\eject

\section{Introduction and summary}
\indent

Studying the scattering properties of black holes is a topic of current and future interest. Of particular importance is the study of the quantum mechanical rate of decay of black holes into a field quantum of a given frequency, which is given, up to a thermal factor, by the greybody factor, due to the effective potential created by the black hole outside its horizon. This greybody factor is, also up to a projection term, related to the cross section of absorption of an equivalent quantum field by the black hole.

The low frequency limit of the absorption cross section for minimally coupled scalar fields is equal to the area of the black hole horizon, a result which can also be extended to higher spin fields \cite{dgm96}. Equivalently, one can say that the low frequency cross section equals four times the Bekenstein--Hawking black hole entropy: $\sigma= 4 G S.$ Recently there has been a renewed interest in this cross section, also from the theoretical point of view, since this quantity is directly related to the shear viscosity $\eta$ of the dual quark-gluon plasma, which according to the fluid--gravity correspondence \cite{Policastro:2001yc,Policastro:2002se} behaves as a strongly coupled (and almost ideal) fluid. In this context it is worth mentioning the KSS bound \cite{Kovtun:2004de}, which states that for theories with a holographic dual the ratio $\eta/s$ between the shear viscosity and the entropy density has a lower bound of $\frac{1}{4\pi}.$ This bound can be saturated for boundary field theories in the limit of large 't Hooft coupling and number of colors.

The KSS bound was established in classical Einstein gravity (without higher order corrections). However, string theories require higher derivative corrections in $\a,$ the inverse string tension.

Such theories are dual to Einstein gravity (without corrections). However, more general computations in higher derivative gravity showed that the KSS bound can be violated, although the crucial sign of the coefficient in front of the higher derivative correction is in general undetermined. This suggests the need to study higher derivative corrections to $\eta,$ and correspondingly to the absorption cross section $\sigma$ \cite{Paulos:2009yk}.

Another motivation to study higher derivative corrections to the absorption cross section is the aforementioned relation $\sigma= 4 G S$ which, like the KSS bound, was only established classically. It is important to check if and how such relation is maintained in the presence of higher derivative terms, namely string $\a$ corrections. These
are a few of the theoretical motivations which lead us to study $\a$ corrections to the absorption cross section. But such study is interesting and important by its own, since gravitational wave astronomy is becoming an experimental reality which could allow for the detection and measurement of (small) string effects.

The first work to discuss the effects of leading $\a$ corrections quadratic in the Riemann tensor in the absorption cross section of spherically symmetric black holes for generic $d$ dimensions was article \cite{Moura:2006pz}, but just dealing with a particular black hole solution. In this article we wish to perform such study for any $d$ dimensional asymptotically flat spherically symmetric black hole with such corrections. We leave the asymptotically de Sitter/Anti de Sitter cases for a future work.

The article is organized as follows. In section 2 we present the generic $\a$--corrected black hole solutions in the background of which we consider a test scalar field. We also discuss the field equation to this scalar field. In section 3 we solve this field equation in different regions of spacetime, using different approximations: close to the horizon, at asymptotic infinity and in the intermediate region. We present solutions, in closed form, for these three regions. After matching these three different solutions, we are able to obtain a general formula for the $\a$--corrected low frequency absorption cross section of the test scalar fields by the black hole. All we have been describing is performed for a generic metric; in section 4, we apply our result to three different known black holes with $\a$ corrections. In section \ref{entropia} we compute the $\a$--corrected entropy of these three black hole solutions, which we compare to the cross sections obtained in section 4. All these results are obtained in a specific scheme (concerning the metric) and based on assumptions about the coordinate system; in section \ref{ecspv}, we obtain expressions for the entropy and cross section which are covariant and independent of metric redefinitions. We end by discussing our results.

\section{Basic setup: $\a$--corrected field equations in $d$ dimensions}

\subsection{Spherically symmetric $\a$--corrected black hole solutions}
\indent

We start by considering a $d$--dimensional string effective action with $\a$ corrections given, in the Einstein scheme \footnote{In this article we adopt the designation ``scheme'' instead of frame, in order to distinguish it from a generic coordinate frame.}, by
\be \label{eef}
\frac{1}{16 \pi G} \int \sqrt{-g} \left( \R - \frac{4}{d-2} \left( \d^\mu \phi \right) \d_\mu \phi + \lambda'\ \mbox{e}^{\frac{4}{d-2} \left( 1 + w \right) \phi} Y(\R) + \mathcal{L}_{\mathrm{matter}}\right) \mbox{d}^dx .
\ee
\noindent
Here, $Y(\R)$ is a scalar polynomial in the Riemann tensor representing the leading higher derivative string corrections to the metric tensor field, and $\lambda'$ is, up to a numerical factor, the suitable power of the inverse string tension $\a$ for $Y(\R)$. The dilaton field is $\phi,$ and $w$ is the conformal weight of $Y(\R)$, with the convention that $w \left( g_{\mu\nu} \right) = +1$ and $w \left( g^{\mu\nu} \right) = -1$. If one is rather just considering higher order gravitational corrections in a non--stringy framework, one can simply take $\phi=0$ in (\ref{eef}): the main results of this article are also valid for such choice, as long as $\lambda'$ is taken as a purely perturbative parameter. $\mathcal{L}_{\mathrm{matter}}$ contains terms, up to the same order in $\a,$ including the metric, the dilaton and also other matter fields depending on the string theory we are considering.

After having eliminated certain terms involving derivatives of $\phi$, which would only contribute at higher orders in our perturbative parameter $\lambda',$ the dilaton and graviton field equations following from the effective action (\ref{eef}) are respectively of the form

\bea
\nabla^2 \phi - \frac{\lambda'}{2}\ \mbox{e}^{\frac{4}{d-2} \left( 1 + w \right) \phi}\ Y(\R) &=& \mathrm{matter\,\, terms}, \label{bdfe} \\
\R_{\mu\nu} + \lambda'\ \mbox{e}^{\frac{4}{d-2} \left( 1 + w \right) \phi} \left( \frac{\delta Y(\R)}{\delta g^{\mu\nu}} + \frac{1}{d-2} Y(\R) g_{\mu\nu} - \frac{1}{d-2} g_{\mu\nu} g^{\rho\sigma} \frac{\delta Y(\R)}{\delta g^{\rho\sigma}} \right) &=& \mathrm{matter\,\, terms}. \label{bgfe}
\eea
The ``matter terms'', coming from $\mathcal{L}_{\mathrm{matter}},$ involve other fields than the metric and the dilaton.

In this article we consider the scattering of massless test scalar fields by a spherically symmetric black hole with string $\a$ corrections in $d$ dimensions. These black holes are solutions to the corrected Einstein equation coming from (\ref{eef}) which are built perturbatively in $\a$ and valid only in regions where $r^2 \gg \a:$ for these black holes the event horizon is much bigger than the string length. They are of the form

\be \label{schwarz}
ds^2 = -f(r)\ dt^2  + g^{-1}(r)\ dr^2 + r^2 d\Omega^2_{d-2}.
\ee

As a solution to the classical Einstein equations, in principle one could take a metric like (\ref{schwarz}), the most general spherically symmetric metric, with two independent functions $f_0(r), g_0(r)$. But at order $\lambda'=0,$ (\ref{bgfe}) reduces to the Einstein equation in vacuum, $\R_{\mu\nu} =0,$ and in this case one can always take $f_0(r)= g_0(r).$

The spherically symmetric solution to the vacuum Einstein equation in $d$ dimensions is the Tangherlini solution \cite{Myers:1986un}, with
\be
f_0(r) =: f_0^T(r) = 1 - \left(\frac{R_H}{r}\right)^{d-3},
\label{tangher} \ee
$R_H$ being the horizon radius.

In order to also include spherically symmetric black holes in the presence of matter, we will allow for a multiplicative factor $c(r):$
\be
f_0(r) = c(r) \left(1 - \left(\frac{R_H}{r}\right)^{d-3}\right).
\label{tangherc} \ee
This will be the form of the function $f_0(r)$ we will be considering. The factor $c(r)$ may encode string effects (see subsection \ref{giveon}); it also allows for charged black holes, which may happen in the presence of gauge fields. The corresponding solution, corresponding to a black hole of mass $M$ and charge $Q,$ is given by \cite{Myers:1986un}
\be
f_0^Q(r) := 1 - \frac{16 \pi M}{(d-2) \Omega_{d-2}}\frac{1}{r^{d-3}} + \frac{2 Q^2}{(d-2) (d-3)}\frac{1}{r^{2(d-3)}};
\label{tangherq} \ee
it can always be reduced to the general form (\ref{tangherc}) by taking
\be
c(r) = \frac{f_0^Q(r)}{f_0^T(r)}.
\label{cextrem} \ee
As it is well known, $f_0^Q(r)$ has in general two simple roots: taking $x=r^{3-d},$ it is a second degree polynomial in $x.$ In this case $R_H$ in (\ref{tangherc}) represents the largest (in $r$) root of $f_0^Q(r).$ If the other root of $f_0^Q(r)$ is $R_-,$ $c(r)$ is therefore proportional to $R_-^{3-d}-x:$ it is a smooth function of $r.$ For the case of an extremal black hole (i.e. when $R_-\equiv R_H:$ $f_0^Q(r)$ has a degenerate double root) we see that $c(R_H) \equiv 0.$

Concerning the $\a$ corrections, we make the general assumption that the functions $f(r), g(r)$ in (\ref{schwarz}) have the form
\be
f(r) = f_0(r) \left(1+ \lambda f_c(r) \right), \, \, g(r)= f_0(r) \left(1+ \lambda g_c(r) \right). \label{fcgc}
\ee
$f_0(r)$ is the classical  solution, while the functions $f_c(r), g_c(r)$ encode the $\a$ higher--derivative perturbative corrections. $\lambda$ is a dimensionless quantity defined in the same way as $\lambda',$ with the same numerical factor but with $\a$ replaced by the dimensionless quotient $\a/R_H^2.$ Here we are assuming the horizon radius $R_H$ itself does not admit any $\a$ corrections; it is always possible to choose a system of coordinates in order for that assumption to be true.

\subsection{The field equation for minimally coupled scalars in the background of spherically symmetric black holes}
\indent

We now consider the low frequency scattering of a massless
minimally coupled test scalar field $\mathcal{H}$ by the black
holes we saw in the previous section. (By ``test'' we mean it does
not affect the evolution of the black hole background.)

First we analyze the case without $\a$ corrections. In this case,
the scalar field obeys the Klein--Gordon equation
\be \label{kg}
\frac{1}{\sqrt{-g}}
\partial_\mu \left[\sqrt{-g} g^{\mu\nu} \partial_\nu \mathcal{H}
\right] =0. \ee

From (\ref{kg}) and the metric (\ref{schwarz}), the scalar field $\mathcal{H}$ obeys therefore a field equation of the type \cite{dgm96}
\be \label{dottigen}
\partial^2_t \mathcal{H} - F^2(r)\ \partial^2_r \mathcal{H} + P(r)\ \partial_r \mathcal{H} + Q(r)\ \mathcal{H} = 0,
\ee
$F(r), P(r), Q(r)$ being functionals of the metric (\ref{schwarz}) and its derivatives, namely of the functions
$f(r), g(r).$

For pure gravity (in the absence of $\a$ corrections) in $d$ dimensions, it is not difficult to obtain such
functionals, which in this case we designate by $F_{\textsf{cl}}, P_{\textsf{cl}}, Q_{\textsf{cl}}$:
\bea
F_{\textsf{cl}} &=& \sqrt{fg}, \nonumber \\
P_{\textsf{cl}} &=& - f \left[ (d-2) \frac{g}{r} + \frac{1}{2}
\left(f'+g'\right) \right], \nonumber \\ Q_{\textsf{cl}} &=& \frac{\ell \left(
\ell + d - 3 \right)}{r^2} f + \frac{(g-f)f'}{r}. \label{fpq0}
\eea

Since we are in a static, spherically symmetric background, the field $\mathcal{H}$ can be redefined and expanded as
\begin{equation}
\Phi(t,r,\theta)= k(r) \mathcal{H}(t,r,\theta)=\sum_{\ell} \Phi_\ell(t,r) Y_{\ell 0...0}(\theta)\,.
\label{sphericalharmonics1}
\end{equation}
where $\ell$ is the angular quantum number associated with the polar angle $\theta$ and $Y_{\ell,\varphi_1,..,\varphi_{d-3}} (\theta)$ are the usual spherical harmonics defined over the $(d-2)$ unit
sphere. $\varphi_1,..,\varphi_{d-3}$ are the azimuthal angles, which in our problem we may set to constants: $\theta$ carries all the angular information. In this case, up to a normalization, $Y_{\ell 0...0}(\theta)$ are the Gegenbauer
polynomials $C_{\ell}^{\frac{d-3}{2}}(\cos\theta)$ \cite{hep-th/0206084}.

It is customary to rewrite the above equation (\ref{dottigen}) in terms of the tortoise coordinate $r_*$ defined in this case by $dr_* = \frac{dr}{F(r)}.$ In order to achieve so, we take in (\ref{sphericalharmonics1})
\be
k(r) = \frac{1}{\sqrt{F}} \exp \left( - \int \, \frac{P}{2F^2} \, dr \right), \label{k}
\ee
\noindent
and replace $\partial/\partial r$ by $\partial/\partial r_*.$

It is then easy to see that an equation like (\ref{dottigen}) may be written as a wave equation with a potential $V \left[ f(r), g(r) \right]$ \cite{dg05a}:

\be
\frac{\partial^2 \Phi}{\partial r_*^2} - \frac{\partial^2 \Phi }{\partial t^2} = \left( Q + \frac{F'^2}{4} - \frac{F F''}{2} - \frac{P'}{2} + \frac{P^2}{4 F^2} + \frac{P F'}{F} \right) \Phi \equiv V \left[ f(r), g(r) \right] \Phi. \label{potential0}
\ee
In the absence of $\a$ corrections, the $\lambda=0$ part of $k(r)$ is given by $k_0(r)=r^{\frac{d-2}{2}}$ (see (\ref{k0}) below). The potential is given by
\bea
V_{\textsf{cl}} \left[ f(r), g(r) \right] &=& Q_{\textsf{cl}} + \frac{F_{\textsf{cl}}'^2}{4} - \frac{F_{\textsf{cl}} F_{\textsf{cl}}''}{2} - \frac{P_{\textsf{cl}}'}{2} + \frac{P_{\textsf{cl}}^2}{4 F_{\textsf{cl}}^2} + \frac{P_{\textsf{cl}} F_{\textsf{cl}}'}{F_{\textsf{cl}}} \nonumber \\
&=& \frac{1}{16 r^2 f g}
\left[(16 \ell (\ell +d-3) f^2 g+ r^2 f^2 f'^2 +3 r^2 g^2 f'^2-2 r^2 f (f+g) f' g' -4 r^2 fg (g-f) f'' \right. \nonumber \\
&+& \left. 16r f g^2 f' +4 r (d-6)f^2 g f' +4 (d-2)r f^2 g g' +4 (d-4) (d-2) f^2 g^2 \right]. \label{potentialcl}
\eea
For solutions (\ref{schwarz}) with $f(r)= g(r)$ a potential analogous to $V_{\textsf{cl}} \left[ f(r), g(r) \right]$ has been obtained in $d$ dimensions in \cite{hep-th/0206084}. Still in the absence of $\a$ corrections, equation (\ref{potential0}) also governs tensor--type gravitational perturbations of the metric, with the same potential $V_{\textsf{cl}} \left[ f(r), g(r) \right]$ obtained from the same functions $F_{\textsf{cl}}(r), P_{\textsf{cl}}(r), Q_{\textsf{cl}}(r)$ in (\ref{fpq0}), as it was shown in \cite{ik03a}.

Determining the field equation for the scalar $\mathcal{H}$ in the presence of the $\a$ corrections is not as simple. Since we are dealing with minimally coupled scalars, one could take the same equation (\ref{potential0}) as in Einstein--Hilbert gravity, and consider that only the graviton is affected by the higher--derivative terms. That procedure is very often followed in the literature, in the context of black holes with non--stringy higher derivative corrections. In such cases, it is possible that the higher derivative corrections only affect the metric. Particularly in the context of Lovelock theories in $d$ dimensions, even the metric does not get a higher order equation of motion, as it is well known.

That cannot be the case in the context of string theory we are considering, since the gravitational correction $Y(\R)$ in (\ref{eef}) is multiplied by a term containing the dilaton and, as we have seen, acts as a source term in its field equation (\ref{bdfe}). In string theory the graviton field equation (\ref{bgfe}) is modified and, in general, it is of higher order. That must be the case of the other field equations too, including the dilaton. The dilaton field equation (\ref{bdfe}) is of second order, but recall that it results from eliminating terms which would be of higher order in $\lambda,$ since $\phi$ is at least of order $\lambda.$ We do not have such information for $\mathcal{H}$, and therefore we cannot make \textit{a priori} a similar elimination, at least without some extra input.

But a clever argument from \cite{Paulos:2009yk} tells us that a second order equation is enough to describe the dynamics of the test scalar field $\mathcal{H}$, at least close to the horizon. Indeed, regularity at the horizon tells us that the scalar field can only depend on one (but not on the two) of the Eddington--Finkelstein coordinates $u=t-r_*,v=t+r_*.$ This means that close to the horizon, $\mathcal{H}$ must satisfy either $\frac{\partial \mathcal{H}}{\partial u}=0$ (incoming solution) or $\frac{\partial \mathcal{H}}{\partial v}=0$ (outgoing solution), which for the metric (\ref{schwarz}) may be written as $\frac{\partial \mathcal{H}}{\partial t} =\pm F(r) \frac{\partial \mathcal{H}}{\partial r}.$ Combining the two possible behaviors results in a second order field equation for $\mathcal{H}$ close to the horizon, of the form
\be
\left(\frac{\partial}{\partial t} - F(r) \frac{\partial}{\partial r}\right) \left(\frac{\partial}{\partial t} + F(r) \frac{\partial}{\partial r}\right)\mathcal{H}=0. \label{eqhor}
\ee
In principle, the $\a$--corrected scalar field equation should also be a higher order differential equation, of the
same order of the derivatives in the corrections we are considering, namely in $Y(\R).$ But, as argued in \cite{Paulos:2009yk}, close to the horizon such higher order equation should reduce to powers (of the same order) of the second order equation (\ref{eqhor}). Therefore in this region one may simply take (\ref{eqhor}).

At infinity, the curvature vanishes for an asymptotically flat solution like those we are considering, and so do the curvature corrections in $Y(\R).$ We assume the same to be true for the $\a$ corrections in general. Therefore the field equation for $\mathcal{H}$ in this region should be the same as if there were no $\a$ corrections, i.e. a second order equation.

As we will see, in order to study the scattering of scalars by black holes we will only need the scalar field equation in the intermediate region between the horizon and asymptotic infinity in order to match the solutions obtained in these two regions. Since in these two regions we have second order scalar field equations, in order to match the respective solutions it is natural to take a second order equation.

This way we assume for the scalar $\mathcal{H}$ a field equation like (\ref{dottigen}), with functionals $F(r), P(r), Q(r)$ of the functions $f(r), g(r),$ but this time including explicit $\lambda$--corrected terms, which we write as:
\be
F = F_{\textsf{cl}}, P = P_{\textsf{cl}} + \lambda P_{\textsf{corr}}, Q = Q_{\textsf{cl}} + \lambda Q_{\textsf{corr}}. \label{fpqc}
\ee
The $\lambda=0$ parts $F_{\textsf{cl}}, P_{\textsf{cl}}, Q_{\textsf{cl}},$ as we have seen, correspond to the $\a=0$ terms for $F, P, Q$ given in (\ref{fpq0}), while $P_{\textsf{corr}}, Q_{\textsf{corr}}$ represent explicit $\lambda$ corrections (with a metric like (\ref{fcgc}) we can always define $F$ as $F_{\textsf{cl}}$ in (\ref{fpq0}), without explicit corrections). Such equation for $\mathcal{H}$ can be rewritten as (\ref{potential0}), but with a potential $V [f(r),g(r)]$ also with explicit $\lambda$ corrections.

Still, as we will see the result for the absorption cross section will be essentially independent of the potential, as long as a few general conditions are respected.

We are now ready to start studying scattering processes in the background of a black hole like (\ref{schwarz}).

\section{Scattering of minimally coupled scalars by spherically symmetric $\a$--corrected non--extremal black holes in $d$ dimensions}
\indent

A classical result in Einstein gravity is that, for any spherically symmetric black hole in arbitrary dimension, the absorption cross section of minimally coupled massless scalar fields is equal to the area of the black hole horizon \cite{dgm96}, or equivalently $\sigma=4 S,$ $S$ being the Bekenstein--Hawking entropy. In order to extend such study to an effective theory with string $\a$ corrections, we shall use the technique of matching solutions, which was first developed for Einstein gravity in $d=4$ in \cite{u76}, and later extended to arbitrary $d$ dimensions in \cite{Harmark:2007jy}. That was also the technique which was used in \cite{Moura:2006pz}, where for the first time black hole scattering with $\R^2$ $\a$ corrections was studied. In that paper, a formula for the absorption cross--section was derived for a particular $d$--dimensional solution \cite{cmp89}. We are looking for a general formula for the absorption cross section, applicable to a general solution like (\ref{schwarz}). The idea of this technique is to separately solve the scalar field equation (\ref{dottigen}) in different regions of the parameter $r,$ where in each region we take a different approximation in order to simplify the equation.

We will be considering scattering at low frequencies, $R_H \omega \ll 1$. The low frequency requirement is necessary in order to use the technique of matching solutions: it is precisely when the wavelength of the scattered field is very large, compared to the radius of the black hole, that one can actually match solutions near the event horizon to solutions at asymptotic infinity \cite{u76, Harmark:2007jy}. Also, at low frequencies, only the mode with lowest angular momentum contributes to the cross section \cite{dgm96}; therefore, from now on, we will always take $\ell=0.$ Since $C_0^{\frac{d-3}{2}}(\cos\theta) \equiv 1,$ from the expansion (\ref{sphericalharmonics1}) we only have to consider $\mathcal{H}_0(t,r) =: H(t,r).$

We assume that the solutions to the field equation (\ref{potential0}) are of the form $\Phi(r_*,t) = e^{i\omega t} \Phi(r_*)$, such that $\frac{\partial\Phi}{\partial t} = i\omega \Phi$ (the same being valid for $H (r, t)$). This way (\ref{potential0}) looks like Schr\"odinger equation.

In appendix \ref{fceqgc} we obtain the temperature $T$ of a black hole solution of the form (\ref{fcgc}): not surprisingly, it is proportional to $f_0'(R_H).$ In this article we assume we are dealing with non--extremal black holes; therefore we take $f_0'(R_H) \neq 0.$ For the same reason, from the discussion following (\ref{cextrem}) and since $f'_0(R_H)= \frac{(d-3) c(R_H)}{R_H},$ we also assume $c(R_H) \neq 0.$ We leave the analysis of scattering by extremal black holes to a future work.

\subsection{Scattering close to the event horizon}
\label{sch}
\indent

We start by solving (\ref{potential0}) near the black hole event horizon. Since $f(R_H) = g(R_H) \equiv 0,$ in this region the functions $f(r), g(r)$ from (\ref{fcgc}) have
the form
\be
f(r) \simeq f'_0(R_H) \left(1+ \lambda f_c(R_H) \right) \left( r-R_H \right), \, \, g(r) \simeq f'_0(R_H) \left(1+ \lambda g_c(R_H) \right) \left( r-R_H \right). \label{fghor}
\ee

We then naturally take the following assumption for the potential $V [f(r),g(r)]$ in (\ref{potential0}): at the horizon it vanishes, and as long as
$\frac{r-R_H}{R_H} \ll \left( R_H \omega \right)^2$ one will have $V [f(r),g(r)] \ll \omega^2$ and in this near--horizon region it may be neglected in (\ref{potential0}).
This assumption is based on the fact that $V [f(r),g(r)]$ is a function of $f(r),g(r),$ which vanish at the horizon. But $V [f(r),g(r)]$ may also be a function of the
derivatives of $f(r),g(r),$ which do not vanish at the horizon. That is the case of the potential $V_{\textsf{T}} [f(r),g(r)]$ as seen from (\ref{potential}). In this case
the combinations of terms including derivatives of $f(r),g(r)$ are such that the assumption is indeed valid. That is necessarily the case for the classical part of
$V [f(r),g(r)]$ (the $\a=0$ part of $V_{\textsf{T}} [f(r),g(r)]$), as it an be seen from (\ref{vthor}); this classical part is universal, and it indeed vanishes at the
horizon. The remaining part of the potential depends on the considered $\lambda$ corrections, and we cannot guarantee that it always vanishes at the horizon, like the
$\a$ correction of $V_{\textsf{T}} [f(r),g(r)]$ in (\ref{vthor}) indeed does. But if that is the case then it is always suppressed by $\lambda,$ which is some power of
$\a,$ guaranteeing the validity of the assumption of the smallness of $V [f(r),g(r)]$ near the horizon.

One thus obtains, very close to the event horizon,
\be
\left( \frac{d^2}{dr_*^2} +\omega^2 \right) \Big( k(r) H (r) \Big) = 0. \label{hhor0}
\ee

In this same region, with $f_0$ given by (\ref{tangherc}), one has from (\ref{k})
\be
k(r)=R_H^{\frac{d-2}{2}} \left[1-\frac{1}{4} \lambda \left(\left(f_c(R_H)+g_c(R_H)\right) - \left(f_c(R_H)-g_c(R_H)\right) \log\left(\frac{r-R_H}{R_H}\right) \right)\right] + {\mathcal{O}} \left( r-R_H \right).
\label{knear}
\ee
One can always choose a scheme in such a way that the condition $f_c(R_H)\equiv g_c(R_H)$ is verified (this assertion  will be clarified in section \ref{dfd}). If this is the case, then $k(r)$ is well defined at $r=R_H,$ and can be treated simply as a constant, $k(R_H),$ in a neighborhood of the horizon. Such constant can be discarded from (\ref{hhor0}), which we may then simply write as
\be
\left( \frac{d^2}{dr_*^2} +\omega^2 \right) H(r) = 0. \label{hhor}
\ee

The solutions to (\ref{hhor}) are plane waves. As we are interested in studying the absorption cross section, we shall consider the general solution for a
purely incoming plane wave:

\be \label{near}
H (r_*) = A_{\text{\tiny{near}}} e^{i \omega r_*} + {\mathcal{O}} \left( r-R_H \right).
\ee

From (\ref{fcgc}) and the definition of $F$ in (\ref{fpqc}), one has $$r_*(r) = \int \frac{1}{f_0(r)} \left(1 - \lambda \frac{f_c(r)+g_c(r)}{2}\right) \,
d\,r.$$
Also with $f_0$ given by (\ref{tangherc}), one has in this region $f_0(r) \simeq f'_0(R_H) \left( r-R_H \right), \, f'_0(R_H)= \frac{(d-3) c(R_H)}{R_H},$
and therefore\footnote{When obtaining (\ref{knear}), we mentioned we have re-scaled the time coordinate in order to have $f_c(R_H)\equiv g_c(R_H).$ This choice could also be obtained by choosing an adequate scheme for the metric (see section \ref{dfd}). As we will see in the same section, our final result is independent of the chosen scheme. We then prefer in general to leave $f_c(R_H)$ and $g_c(R_H)$ as independent quantities.}
\be
r_*(r) = \frac{R_H}{(d-3) c(R_H)} \left( 1 - \lambda \frac{f_c(R_H)+g_c(R_H)}{2}\right) \log \left( \frac{r-R_H}{R_H} \right) + {\mathcal{O}} \left( r-R_H
\right). \label{expa}
\ee
Replacing (\ref{expa}) in (\ref{near}), one finally obtains in this region
\be \label{close}
H (r) \simeq A_{\text{\tiny{near}}} \left( 1 + i\frac{R_H \omega}{(d-3) c(R_H)} \left( 1 - \lambda \frac{f_c(R_H)+g_c(R_H)}{2} \right) \log \left(
\frac{r-R_H}{R_H} \right) \right) + {\mathcal{O}} \left( r-R_H \right).
\ee

\subsection{Scattering at asymptotic infinity}
\label{sai}
\indent

We now analyze the solution to (\ref{potential0}) close to infinity.

In this article we consider asymptotically flat black holes which, at infinity, behave like flat Minkowski spacetime. This is equivalent to saying that, in the metric (\ref{schwarz}), functions $f(r), g(r)$ tend to the constant value 1 in the limit of very large $r,$ and their derivatives tend to 0 in the same limit. This means that, as $r \rightarrow \infty,$ $c(r)$ in (\ref{tangherc}) must go to 1; that is the case, for instance, of (\ref{cextrem}). In the same limit $r \rightarrow \infty,$ $f_c(r), g_c(r)$ in (\ref{fcgc}) must go to 0; indeed, in an asymptotically flat space the curvature tensor vanishes at infinity and so should its effects. Therefore in this limit we only need to consider the classical potential $V_{\textsf{cl}} (r),$ without string corrections, like in \cite{Harmark:2007jy}. Having all this in mind, from (\ref{potentialcl}) we obtain that, asymptotically, $V_{\textsf{cl}} (r) \approx \frac{(d-4)(d-2)}{4 r^2} + {\mathcal{O}} \left(\frac{R_H}{r^3}\right),$ and therefore the potential can be neglected in the limit $r \rightarrow \infty.$

This way, in this limit (\ref{potential0}) reduces to a simple free--field equation whose solutions are either incoming or outgoing plane--waves in the tortoise coordinate. One can also solve the same equation in the original radial coordinate in terms of Bessel functions, obtaining \cite{dgm96, u76, Harmark:2007jy} $$H (r) = \left( r \omega \right)^{(3-d)/2} \left[ A_{\text{\tiny{asymp}}}\, J_{(d-3)/2} \left( r\omega \right) + B_{\text{\tiny{asymp}}}\, N_{(d-3)/2} \left( r\omega \right) \right].$$ At low--frequencies, with $r\omega \ll 1$, such solution becomes

\be \label{far}
H (r) \simeq A_{\text{\tiny{asymp}}}\ \frac{1}{2^{\frac{d-3}{2}} \Gamma \left( \frac{d-1}{2} \right)} + B_{\text{\tiny{asymp}}}\ \frac{2^{\frac{d-3}{2}} \Gamma \left( \frac{d-3}{2} \right)}{\pi \left( r\omega \right)^{d-3}} + {\mathcal{O}} \left( r\omega \right).
\ee

\noindent
In order to compute the absorption cross section, we will need to relate the coefficients $A_{\text{\tiny{asymp}}}$ and $B_{\text{\tiny{asymp}}}$ to $A_{\text{\tiny{near}}}$, obtained in (\ref{close}). This can be done by the technique of matching near--horizon to asymptotic solutions, and requires us to solve the scalar field equation (\ref{dottigen}) in an intermediate region, between the event horizon and asymptotic infinity \cite{u76, Harmark:2007jy}. This is what we will do in the following.

\subsection{Scattering in the intermediate region}
\label{sir}
\indent

We now consider the intermediate region: far from the horizon, but not asymptotic infinity. We keep working in the low frequency regime, but this time without any restrictions to the magnitude of the potential, which may be large (but always assumed to be regular).

We want to solve (\ref{potential0}) or, equivalently, equation (\ref{dottigen}), perturbatively in $\lambda.$ We then define the expansion
$$H (r) = H_0 (r) + \lambda H_1 (r), k (r) = k_0 (r) + \lambda k_1 (r).$$
Using the previous assumptions (\ref{fcgc}) for $f, g$ and taking their $\lambda=0$ term (which is $f_0$), and also using the $\lambda=0$ terms $F_{\textsf{cl}}, P_{\textsf{cl}}, Q_{\textsf{cl}}$ from (\ref{fpq0}) in (\ref{dottigen}), we obtain the following equation for $H_0 (r),$ written in the $r$ coordinate (where if $\lambda=0$ $\frac{d}{dr_*}=f_0 \frac{d}{dr}$):

\be \label{h0}
\left[ - f_0(r) \frac{d}{dr} \left( f_0(r) \frac{d}{dr} \right) + f_0(r) \left( \frac{(d-2) (d-4) f_0(r)}{4 r^2} + \frac{(d-2) f_0'(r)}{2r} \right) \right] \Big( k_0 (r) H_0 (r) \Big) = 0,
\ee

First, one verifies that from (\ref{fpq0}) and (\ref{k}) we have, up to a multiplicative constant (and for any $f$),
\be \label{k0}
k_0(r) = \frac{1}{\sqrt{f}} \exp \left( \int \, \left(\frac{d-2}{2r}+\frac{f'}{2f}\right) \, dr \right) = r^{\frac{d-2}{2}}.
\ee
Replacing this expression for $k_0,$ one indeed has, after a simple computation,
\be
\frac{d}{dr} \left( f_0(r) \frac{d}{dr} \left(r^{\frac{d-2}{2}} H_0 \right) \right)= \left( \frac{(d-2) (d-4) f_0(r)}{4 r^2} + \frac{(d-2) f_0'(r)}{2r} \right) r^{\frac{d-2}{2}} H_0 + r^{\frac{2-d}{2}} \frac{d}{dr} \left( r^{d-2} f_0(r) \frac{d}{dr}  H_0 \right).
\ee
Replacing (\ref{h0}) in the equation above, we see that $H_0 (r)$ satisfies
\be \label{h0simp}
\frac{d}{dr} \left( r^{d-2} f_0(r) \frac{d}{dr} H_0 (r)\right)  = 0,
\ee
whose most general solution is\footnote{The integrals in this subsection are all meant to be indefinite.}

\be \label{h0sol}
H_0 (r) = A_{\text{\tiny{inter}}}^0 + B_{\text{\tiny{inter}}}^0 \int\frac{d\,r}{r^{d-2} f_0(r)}.
\ee

In order to solve for $H_1 (r),$ we take for $F, P, Q$ similar expansions as we did for $H, \, k:$ $F = F_0 + \lambda F_1, P = P_0 + \lambda P_1, Q = Q_0 + \lambda Q_1.$ The $\lambda=0$ parts $F_0, P_0, Q_0,$ as we have seen, correspond to the $\a=0$ terms for $F, P, Q$ given in (\ref{fpq0}), while $F_1, P_1, Q_1$ represent the full $\lambda$ corrections: those which are explicit, from (\ref{fpqc}), and those which are implicit, coming by replacing the $\lambda$ corrections (\ref{fcgc}) to $f, g$ in (\ref{fpq0}).

We then expand every term of (\ref{dottigen}). To zero order in $\lambda$ we obtain
\be \label{h0mod}
H_0'' - \frac{P_0}{F_0^2} H_0' - \frac{Q_0}{F_0^2} H_0 =0,
\ee
which is completely equivalent to (\ref{h0}), with solution (\ref{h0sol}).

The terms of first order in $\lambda$ are $-F_0^2 H_1'' -F_1^2 H_0''+P_0 H_1' +P_1 H_0' +Q_0 H_1 +Q_1 H_0,$ which may be rewritten as
\be \label{h1}
H_1'' - \frac{P_0}{F_0^2} H_1' - \frac{Q_0}{F_0^2} H_1= R(r), \,\, R(r)=-\left(\frac{F_1}{F_0}\right)^2 H_0'' + \frac{P_1}{F_0^2} H_0' + \frac{Q_1}{F_0^2} H_0
\ee
This is a second--order linear nonhomogeneous differential equation for $H_1.$ The homogeneous part is exactly the same as the differential equation (\ref{h0mod}) for $H_0,$ with general solution (\ref{h0sol}), replacing $H_0 (r), A_{\text{\tiny{inter}}}^0, B_{\text{\tiny{inter}}}^0 $ by $H_1 (r), A_{\text{\tiny{inter}}}^1, B_{\text{\tiny{inter}}}^1 .$ A basis for the vector space of independent solutions of (\ref{h0mod}) is
\be
h_1(r)=1, \,\, h_2(r)=\int\frac{d\,r}{r^{d-2} f_0(r)}; \label{h1h2}
\ee
the respective wronskian matrix is
$$W(r)=\left[ \begin{array}{cc}
               h_1(r) & h_2(r) \\
               h'_1(r) & h'_2(r) \\
             \end{array}
           \right]
           = \left[
             \begin{array}{cc}
              1 & \int\frac{d\,r}{r^{d-2} f_0(r)}\\
              0 & \frac{1}{r^{d-2} f_0(r)}\\
             \end{array}
           \right],$$
with inverse
$$W^{-1}(r)=r^{d-2} f_0(r) \left[
             \begin{array}{cc}
              \frac{1}{r^{d-2} f_0(r)} & -\int\frac{d\,r}{r^{d-2} f_0(r)}\\
              0 & 1\\
             \end{array}
           \right].$$
According to the method of variation of constants, a particular solution to the nonhomogeneous equation (\ref{h1}) is given by
\be \label{h1part}
H_1^{\text{\tiny{part}}} (r) = v_1(r) h_1 (r) + v_2(r) h_2 (r), \,\,
\left[
    \begin{array}{c}
      v_1(r) \\
      v_2(r) \\
    \end{array}
  \right]= \int R(r) \, W^{-1}(r) \,
\left[
    \begin{array}{c}
      0 \\
      1 \\
    \end{array}
  \right] \, d\,r.
\ee
To obtain the most general solution to (\ref{h1}) one just needs to add to $H_1^{\text{\tiny{part}}} (r)$ the most general solution (\ref{h0sol}) to the homogeneous equation (\ref{h0mod}), including the contributions $H_0, H_1$ as $H=H_0 + \lambda H_1:$

\be
H (r) = \left(A_{\text{\tiny{inter}}}+ \lambda v_1(r) \right)h_1 (r) + \left(B_{\text{\tiny{inter}}} + \lambda v_2(r)\right) h_2 (r) = A_{\text{\tiny{inter}}} + B_{\text{\tiny{inter}}} \int\frac{d\,r}{r^{d-2} f_0(r)} + \lambda H_1^{\text{\tiny{part}}} (r). \label{h1mod}
\ee
We still need to verify the behavior of the function $H_1^{\text{\tiny{part}}} (r)$ given by (\ref{h1part}), namely of the indefinite integrals
\be
v_1(r)= - \int R(r) r^{d-2} f_0(r) h_2 (r) \, d\,r, \,\, v_2(r)= \int R(r) r^{d-2} f_0(r) \, d\,r. \label{v1v2}
\ee
Since the metric, by assumption, has no other singularity than the horizon, $v_1(r), \, v_2(r)$ should be well defined functions for $r > R_H.$ It is therefore necessary to verify that the integrals at $v_1(r), \, v_2(r)$ converge at infinity (i.e for arbitrarily large values of $r$) and to study their behavior close to $r=R_H.$

Close to infinity, one has at most
\be
f_0(r) = 1 - \left(\frac{m}{r}\right)^{d-3} + {\mathcal{O}} \left(\frac{m}{r}\right)^{d-2}; \label{f0inf}
\ee
no lower power of $\frac{1}{r}$ is allowed \cite{Myers:1986un}. (If $f_0(r)$ represents the Tangherlini solution (\ref{tangher}), then $m$ represents the horizon radius $R_H,$ but in general other string effects may be present.) Taking this to be the asymptotic form of $f_0(r),$ one has
\bea
v_1(r) &\approx& - \int \frac{d-3}{2} B_{\text{\tiny{inter}}} \frac{m^{d-5}}{r^{d-3}} \left(f_c'(r)+ g_c'(r)\right) \, d\,r, \nonumber \\
v_2(r) &\approx& \frac{d-3}{2} B_{\text{\tiny{inter}}} \left(f_c(r)+ g_c(r)\right). \label{v1v2inf}
\eea
Here we make the same assumptions as in section \ref{sai}, namely that all the $\lambda$--corrections (the functions $f_c(r), g_c(r)$ and the functionals $P_{\textsf{corr}}, Q_{\textsf{corr}}$) tend to zero at infinity (i.e. asymptotically the effects of the corrections vanishes, and everything happens as if $f=g=f_0$). This is a very reasonable physical assumption. In such case, $v_1(r), \, v_2(r)$ and therefore $H_1^{\text{\tiny{part}}} (r)$ vanish at infinity.

Close to the horizon, one has
\bea
v_1(r) &\approx& \frac{B_{\text{\tiny{inter}}}}{4 (d-3) c(R_H) R_H^{d-1}} \left(f_c(R_H)-g_c(R_H)\right) \log\left(\frac{r-R_H}{R_H}\right)^2 +v^{\text{\tiny{reg}}}_1(R_H)+ {\mathcal{O}} \left(\frac{r-R_H}{R_H}\right), \nonumber \\
v_2(r) &\approx& - \frac{B_{\text{\tiny{inter}}}}{2 R_H^2} \left(f_c(R_H)-g_c(R_H)\right) \log\left(\frac{r-R_H}{R_H}\right) +v^{\text{\tiny{reg}}}_2(R_H) + {\mathcal{O}} \left(\frac{r-R_H}{R_H}\right). \label{v1v2hor}
\eea
$v^{\text{\tiny{reg}}}_1(R_H), v^{\text{\tiny{reg}}}_2(R_H)$ are defined up to two integration constants (from (\ref{v1v2})), which may be absorbed by $A_{\text{\tiny{inter}}}$ in (\ref{h1mod}). The only terms in these functions which are not regular at $r=R_H$ are both multiplied by $\left(f_c(R_H)-g_c(R_H)\right)$ but, as we have mentioned in section \ref{sch} and will clarify in section \ref{dfd}, we can always choose a scheme in order to obtain a system of coordinates such that $f_c(R_H)\equiv g_c(R_H)$. The remaining terms in $v_1(r), v_2(r)$  are regular and vanish at $r=R_H.$ This means that one can ignore $H_1^{\text{\tiny{part}}} (r)$ close to the horizon and simply consider the solution to the homogeneous equation.

To summarize: we were able to solve the field equation (\ref{dottigen}) equation in the intermediate region. This is a linear nonhomogeneous equation; for its general solution, we should add to the solution to the homogeneous equation a particular solution $H_1^{\text{\tiny{part}}} (r),$ which we found by the method of variation of constants. We verified the behavior of this particular solution $H_1^{\text{\tiny{part}}} (r)$ at infinity and close to the black hole horizon, and in both cases we concluded that either it vanishes or its contribution was subleading; close to these regions, we can neglect $H_1^{\text{\tiny{part}}} (r)$ and simply consider the solution to the homogeneous equation $H_0(r).$ This will be a key feature for the matching process.

\subsection{Calculation of the absorption cross section}
\label{cacs}
\indent

We are now ready to start the matching process, using $f_0$ given by (\ref{tangherc}).

If we evaluate (\ref{h1mod}) near the horizon, from (\ref{h1h2}) we obtain
\be
H (r) \simeq A_{\text{\tiny{inter}}} + \frac{B_{\text{\tiny{inter}}}}{(d-3) R_H^{d-3} c\left(R_H\right)} \log \left( \frac{r-R_H}{R_H} \right) + {\mathcal{O}} \left(\frac{r-R_H}{R_H}\right).
\ee

Matching the coefficients above to the ones in (\ref{close}) immediately yields
\bea
A_{\text{\tiny{near}}} &=& A_{\text{\tiny{inter}}}, \nonumber \\
B_{\text{\tiny{inter}}} &=& i A_{\text{\tiny{near}}} R_H^{d-2} \omega \left( 1 - \lambda \frac{f_c(R_H)+g_c(R_H)}{2}\right).
\eea

As in section \ref{sai} we assume that $c(r) \underset{r \rightarrow \infty}{\longrightarrow} 1$ in such a way that condition (\ref{f0inf}) is verified. This condition allows us to have, at asymptotic infinity, to leading order,
\be
h_2(r) =\int\frac{d\,r}{r^{d-2} f_0(r)} \simeq \int\frac{d\,r}{r^{d-2} } + \cdots =
-\frac{1}{d-3} \frac{1}{r^{d-3}} + \cdots,
\ee
and, therefore, evaluating (\ref{h1mod}) again asymptotically,
\be
H (r) \simeq A_{\text{\tiny{inter}}} - \frac{B_{\text{\tiny{inter}}}}{d-3} \frac{1}{r^{d-3}} + \cdots.
\ee

\noindent
In this region one may match the coefficients above to the ones in (\ref{far}), yielding

\bea
A_{\text{\tiny{asymp}}} &=& 2^{\frac{d-3}{2}} \Gamma \left( \frac{d-1}{2} \right) A_{\text{\tiny{inter}}} = 2^{\frac{d-3}{2}} \Gamma \left( \frac{d-1}{2} \right) A_{\text{\tiny{near}}}, \nonumber \\
B_{\text{\tiny{asymp}}} &=& - \frac{\pi \omega^{d-3}}{2^{\frac{d-3}{2}} (d-3) \Gamma \left( \frac{d-3}{2} \right)} B_{\text{\tiny{inter}}} = - \frac{i \pi \left( R_H \omega \right)^{d-2}}{2^{\frac{d-1}{2}} \Gamma \left( \frac{d-1}{2} \right)} \left( 1 - \lambda \frac{f_c(R_H)+g_c(R_H)}{2}\right) A_{\text{\tiny{near}}}. \label{match}
\eea

Computing the low frequency absorption cross section is now a simple exercise in scattering theory \cite{u76, Harmark:2007jy}. Near the black hole event horizon, from (\ref{near}), the incoming flux per unit area is
\be \label{jnear}
J_{\text{\tiny{near}}} = \frac{1}{2i} \left( H^\dagger (r_*) \frac{d H}{d r_*} - H (r_*) \frac{d H^\dagger}{d r_*} \right) = \omega \left| A_{\text{\tiny{near}}} \right|^2.
\ee
The outgoing flux per unit area at asymptotic infinity, where $r_*$ and $r$ coincide, is, from (\ref{far}),
\be
J_{\text{\tiny{asymp}}} = \frac{1}{2i} \left( H^\dagger (r) \frac{d H}{dr} - H (r)
\frac{d H^\dagger}{dr} \right) =\frac{2}{\pi} r^{2-d} \omega^{3-d} \left| A_{\text{\tiny{asymp}}}
B_{\text{\tiny{asymp}}} \right|.
\ee
In order to compute the cross section, this same flux per unit area at asymptotic infinity must be integrated over a sphere of (large) radius $r$, and the result should be divided by the incoming flux per unit area:
\be
\sigma = \frac{\int r^{d-2} J_{\text{\tiny{asymp}}}d \Omega_{d-2}}{J_{\text{\tiny{near}}}}= \frac{2}{\pi}  \omega^{2-d} \frac{\left| A_{\text{\tiny{asymp}}}
B_{\text{\tiny{asymp}}} \right|}{\left| A_{\text{\tiny{near}}} \right|^2} \Omega_{d-2}. \label{jj}
\ee
Replacing the results from (\ref{match}), the final result is

\be
\sigma = A_H \left(1 - \lambda \ \frac{f_c(R_H) + g_c(R_H)}{2} \right), \label{seccao}
\ee
where $A_H=R_H^{d-2} \Omega_{d-2}$ is the horizon area with respect to the metric induced by (\ref{schwarz}).

\subsection{Discussion on dependence under field redefinitions}
\label{dfd}
\indent

During our calculation process, we have made the assumption that $f_c(R_H) = g_c(R_H).$ In general, $f_c(r)$ and $g_c(r)$ in (\ref{fcgc}) are two independent functions. Setting them equal by a conformal transformation is possible: that would be equivalent to setting the functions $f(r), g(r)$ in the metric (\ref{schwarz}) equal. That field redefinition is called a change of scheme. By requiring that $f_c(r)=g_c(r)$, we are therefore picking a particular scheme, since such relation is not valid in every scheme.

One may therefore ask which quantities depend and which do not on the choice of scheme. It turns out that physical quantities should not depend on such choice, since schemes are all equivalent up to metric redefinitions.

In appendix \ref{fceqgc} we obtain the temperature $T$ of a black hole solution of the form (\ref{fcgc}), given by eq. (\ref{temp}). With $f_0$ given by (\ref{tangherc}), this temperature comes as
\be
T=\frac{(d-3) c(R_H)}{4 \pi R_H} \left(1+ \lambda \delta T \right). \label{temph}
\ee
In this expression, $c(R_H)$ should only depend on the black hole mass and charges. One can always choose a system of coordinates (namely, rescaling the time coordinate, and its periodicity $1/T$) in which $T$ has units such that we have $c(R_H) \equiv 1.$

$R_H$ is the horizon radius in the scheme one is considering. Of course the horizon location does not depend on the scheme, but if one wants the metric to remain of the form (\ref{schwarz}) (as we do), after the change of scheme one must apply a change of coordinates. The relation between the location of the horizon in the two different coordinates can be obtained precisely by equating the expressions for the temperature in the two different schemes, since the black hole temperature, as a physical quantity, does not depend on the chosen coordinates or schemes. Consider for example the known Einstein and string schemes, with horizon radii $R_H^E, R_H^S$ and $\lambda$ corrections $\delta T_E, \delta T_S,$ respectively. Writing (\ref{temph}) in terms of the variables of each of the two schemes and equating the corresponding expressions, one obtains the desired relation between the horizon locations in the two different schemes:
\be
R_H^E = R_H^S \left(1+ \lambda \left(\delta T_E - \delta T_S \right) \right). \label{rhes}
\ee

In general, for a black hole of the type we have been considering, its mass can be written with a perturbative multiplicative $\lambda$--correction to the classical Tangherlini mass ($\Omega_{d-2}=\frac{2 \pi^{\frac{d-1}{2}}}{\Gamma\left(\frac{d-1}{2}\right)}$):
\be
M= \left( 1 + \lambda \ \delta M \right) \frac{\left(d-2\right) \Omega_{d-2}}{16 \pi G} R_H^{d-3}. \label{massa}
\ee
Also in this expression, $R_H$ is the horizon radius and $\delta M$ the $\lambda$ correction in the scheme one is considering. Since the black hole mass also does not depend on the choice of coordinates or schemes, by expressing (\ref{massa}) in the Einstein and string schemes and equating the two expressions, like we did with the temperature, we can reobtain the relation (\ref{rhes}) between $R_H^E$ and $R_H^S$, this time given in terms of the $\lambda$ corrections $\delta M_E, \delta M_S$ respectively in the Einstein and string schemes:
\be
R_H^E = R_H^S \left(1+ \lambda \frac{\delta M_S - \delta M_E}{d-3} \right). \label{rhesm}
\ee
This relation between $R_H^E$ and $R_H^S$ must be unique; therefore, there must be a relation between the mass and temperature $\lambda$ corrections such that (\ref{rhes}) and (\ref{rhesm}) represent exactly the same expression.

The $\lambda$ corrections we have been considering are multiplicative; also for the black hole absorption cross section the result (\ref{seccao}) we obtained is of the form $\sigma=\left.\sigma\right|_{\a=0} \left(1+ \lambda \delta \sigma \right).$ Here $\delta \sigma$, like $\delta M$ and $\delta T,$ is a dimensionless factor characteristic to the specific solution one is considering. These factors also depend on the scheme one is using, as we saw.

Comparing (\ref{seccao}) with (\ref{temp}), we see that
\be
\sigma = A_H \left(1 - \lambda \ \delta T \right), \, \delta T=-\delta \sigma =\frac{f_c(R_H) + g_c(R_H)}{2}.
\label{deltat}
\ee
This relation between $\delta \sigma$ and $\delta T$ will help us expressing the cross section in a way that is independent of the chosen scheme. Once that is achieved, one can simply obtain $\delta \sigma$ by computing the black hole temperature, without having to be concerned with choosing a scheme such that $f_c(R_H) \equiv g_c(R_H).$ We will return to this subject in section \ref{ecspv}.

\section{Application to concrete string--corrected black hole solutions}
\indent

We now apply our results to the computation of the absorption cross section for a few specific black hole solutions in string theory. Although our results can of course be applied to concrete solutions in specific given $d$ dimensions, we prefer to consider in this article only solutions in which $d$ remains arbitrary.

In this article we only consider solutions with leading corrections quadratic in the Riemann tensor. Concretely, we take $Y(\R)= \frac{1}{2}\ \R^{\mu\nu\rho\sigma} \R_{\mu\nu\rho\sigma}$ in (\ref{eef})\footnote{Any other gravitational correction of the same order in $\a$ is equivalent to this one by field redefinitions \cite{cmp89}.}, with $\lambda' = \frac{\a}{2}, \frac{\a}{4}$ or $0$ (or $\lambda = \frac{\a}{2 R_H^2}, \frac{\a}{4 R_H^2}$ or $0$) for bosonic, heterotic and type II \footnote{Type II supersymmetry prevents this term to appear in the ten dimensional effective action; this is why in this case we have $\lambda', \lambda =0$.} strings, respectively. Since in the previous sections we took $\lambda$ with arbitrary higher order corrections and here we are working with a specific correction of first order in $\a,$ we prefer in this section to keep $\a$ explicit. We choose to work in the context of heterotic strings and, therefore, we take $\lambda = \frac{\a}{4 R_H^2}$ in this section. The formulas we obtain in this section are also valid in the context of bosonic strings by simply replacing $\a$ by $2 \a.$

\subsection{The $d$--dimensional Callan--Myers--Perry black hole}
\indent

The Callan--Myers--Perry solution was the first $d$--dimensional black hole solution with quadratic Riemann corrections to be obtained (in \cite{cmp89}). It is a simple generalization of the Tangherlini solution of the form (\ref{fcgc}), with $f_0=f_0^T$ given by (\ref{tangher}) and (in the Einstein scheme)
\be
f_c(r)=g_c(r)=f_c^{CMP}(r):=- \frac{(d-3)(d-4)}{2}\ \left(\frac{R_H}{r}\right)^{d-3}\ \frac{1 - \left( \frac{R_H}{r}\right)^{d-1}}{1 - \left(\frac{R_H}{r}\right)^{d-3}}. \label{fccmp}
\ee
A simple application of l'H\^opital's rule allows us to compute
\be
\lim_{r \to R_H}{f_c^{CMP}(r)}=f_c^{CMP}(R_H) =-\frac{(d-1) (d-4)}{2}, \label{hopital}
\ee
from which, using (\ref{temp}), we obtain the temperature of the Callan--Myers--Perry black hole in the Einstein scheme:
\be
T=\frac{d-3}{4 \pi R_H^E} \left(1 + \delta T_E^{CMP}\ \frac{\a}{4 \left(R_H^E\right)^2} \right), \ \delta T_E^{CMP} = - \frac{(d-1) (d-4)}{2}. \label{tcmpe}
\ee
From (\ref{deltat}), we obtain the absorption cross section in the Einstein scheme \footnote{Here we are just confirming the result of \cite{Moura:2006pz}, where this same computation was performed, with less generality, just for this particular solution.}
\be
\sigma = A_H^E \left( 1 + \frac{(d-1) (d-4)}{8}\ \frac{\a}{\left(R_H^E\right)^2} \right), \label{seccaocmp}
\ee
with $A_H^E=\left(R_H^E\right)^{d-2} \Omega_{d-2}.$ Just for future reference, the black hole mass is given in this case by
\be
M= \left( 1 + \delta M_E^{CMP}\ \frac{\a}{4 \left(R_H^E\right)^2}\right) \frac{\left(d-2\right) \Omega_{d-2}}{16 \pi G} \left(R_H^E\right)^{d-3}, \ \delta M_E^{CMP} = \frac{(d-3) (d-4)}{2}. \label{mcmpe}
\ee

Also for future reference, in the string scheme \cite{cmp89} the Callan--Myers--Perry solution is still of the form (\ref{schwarz}), but with $f, g$ replaced by $f^{CMP}_S, g^{CMP}_S,$ given by
\bea
f^{CMP}_S(r)&=&f_0^T \left(1+\frac{\a}{2 \left(R_H^S\right)^2} \mu(r)\right), \label{fscmp} \\
g^{CMP}_S(r)&=&f_0^T \left( 1-\frac{\a}{2 \left(R_H^S\right)^2} \epsilon(r)\right), \label{gscmp}
\eea
with the definitions \footnote{The digamma function is given by
$\psi(z)=\Gamma'(z)/\Gamma(z),$ $\Gamma(z)$ being the usual
$\Gamma$ function. For positive $n,$ one defines
$\psi^{(n)}(z)=d^n\,\psi(z)/d\,z^n.$ This definition can be
extended for other values of $n$ by fractional calculus analytic
continuation. These are meromorphic functions of $z$ with no
branch cut discontinuities.

$\gamma$ is Euler's constant, defined by $\gamma=\lim_{n \to
\infty}{\left(\sum_{k=1}^n \frac{1}{k} - \ln n \right)},$ with
numerical value $\gamma \approx 0.577216.$}
\bea
\epsilon(r)&=&\frac{d-3}{4} \frac{\left(\frac{R_H}{r}\right)^{d-3}}{1-\left(\frac{R_H}{r}\right)^{d-3}} \left[\frac{(d-2)(d-3)}{2}
-\frac{2(2d-3)}{d-1} + (d-2)\left(\psi^{(0)}\left(\frac{2}{d-3}\right) + \gamma \right)\right. \nonumber \\ &+& \left. d \left(\frac{R_H}{r}\right)^{d-1} +\frac{4}{d-2} \varphi(r) \right], \label{epsilon} \\
\mu(r)&=&-\epsilon(r)+\frac{2}{d-2} (\varphi(r)-r \varphi'(r)), \label{mu} \\
\varphi(r)&=& \frac{(d-2)^2}{4}
\left[\ln\left( 1 - \left(\frac{R_H}{r}\right)^{d-3} \right) -\frac{d-3}{2}\left(\frac{R_H}{r}\right)^2 - \frac{d-3}{d-1}
\left(\frac{R_H}{r}\right)^{d-1} \right. \nonumber \\
&+& \left. B \left(\left(\frac{R_H}{r}\right)^{d-3};\, \frac{2}{d-3}, 0 \right)\right], \label{fr2} \\
\varphi'\left(r\right)&=&\frac{(d-3) (d-2)^2}{4} \frac{R_H^{d-3}}{r^{d-2}} \frac{1-\left(\frac{R_H}{r}\right)^{d-1}}{1-\left(\frac{R_H}{r}\right)^{d-3}}  \label{filinha}.
\eea
$f_0^T$ is given by (\ref{tangher}), but in the string scheme, the same being valid for $\varphi(r)$ in (\ref{fr2}): in both cases with $R_H$ replaced by $R_H^S$. $B(x;\,a,b)=\int_0^x t^{a-1}\,(1-t)^{b-1}\,dt \!$ is the incomplete Euler beta function.

At the horizon, we have \cite{Moura:2009it}
\be
\varphi\left(R_H\right)=- \frac{(d-2)^2}{8
(d-1)} \left(d^2-2d+2 (d-1) \left(\psi^{(0)}\left(\frac{2}{d-3}\right) + \gamma \right) -3\right). \label{firh}
\ee

In such scheme and system of coordinates, after determining the limits of $\epsilon(r), \mu(r)$ when $r\rightarrow R_H$ (using the definitions (\ref{epsilon}), (\ref{mu}) but also the properties (\ref{hopital}), (\ref{filinha}), (\ref{firh})), from (\ref{temp}) the black hole temperature is given by
\bea
T&=&\frac{d-3}{4 \pi R_H^S} \left(1 + \delta T_S^{CMP} \ \frac{\a}{4 \left(R_H^S\right)^2} \right), \nonumber \\
\delta T_S^{CMP} &=&- \frac{3 d(d-3)\left(d-\frac{5}{3}\right)- 2(d-1)^2 +2 (d-2)(d-1) \left(\psi^{(0)}\left(\frac{2}{d-3}\right) + \gamma \right)}{4(d-1)}. \label{tcmps}
\eea
The black hole mass is given, again in the string scheme, by
\be
M= \left( 1 + \delta M_S^{CMP}\ \frac{\a}{4 \left(R_H^S\right)^2}\right) \frac{\left(d-2\right) \Omega_{d-2}}{16 \pi G} \left(R_H^S\right)^{d-3}, \ \delta M_S^{CMP} = (d-3)\left(-\delta T_S^{CMP}-\frac{(d-2)(d-4)}{2}\right). \label{mcmps}
\ee

\subsection{The string--corrected dilatonic $d$--dimensional black hole}
\indent

The Callan--Myers--Perry solution expresses the effect of the string $\a$ corrections, but it does not consider any other string effects, namely the fact that string theories live in $d_S$ spacetime dimensons ($d_S=10$ or 26 on heterotic or bosonic strings, respectively), and have to be compactified to $d$ dimensions on a $d_S - d$--dimensional manifold. When passing from the string to the Einstein scheme, the volume of the compactification manifold becomes spatially varying. In the simple case when such manifold is a flat torus, that volume depends only on the $d-$dimensional part of the dilaton $\phi$ and, after solving the $\a$--corrected field equation (\ref{bgfe}) the metrics of the compactification manifold and of the $d$--dimensional spacetime decouple.

The explicit solution was worked out in \cite{Moura:2009it}. The general solution for the dilaton, in the background of the spherically symmetric Tangherlini black hole (\ref{tangher}), is necessarily of order $\a:$ $\phi(r) := \frac{\a}{4 R_H^2} \varphi(r)$, with $\varphi(r)$ given by (\ref{fr2}). The derivative of $\phi$ can be obtained from (\ref{filinha}), which can also be written as $r \varphi' =-\frac{(d-2)^2}{2(d-4)} f_c^{CMP}(r),$ with $f_c^{CMP}(r)$ given by (\ref{fccmp}).

The $d$--dimensional part of the metric is of the form (\ref{schwarz}), with $f, g$ given by (\ref{fcgc}), $f_0=f_0^T$ given by (\ref{tangher}) and (in the Einstein scheme)
\be
g_c(r) = f_c^{CMP}(r),\, \, f_c(r)= f_c^{CMP}(r) + 4 \frac{d_S-d}{\left(d_S-2\right)^2} \left(\varphi - r \varphi'\right). \label{fcdil}
\ee
Using (\ref{hopital}), (\ref{filinha}) and (\ref{firh}), one can determine $\lim_{r \to R_H}{\left(\varphi - r \varphi'\right)},$ which is a finite quantity. Together with (\ref{fcdil}) and  again (\ref{hopital}), this allows us to obtain, using (\ref{temp}), the black hole temperature in the Einstein scheme:
\bea
T&=&\frac{d-3}{4 \pi R_H^E} \left(1 + \delta T_E^d\ \frac{\a}{4 \left(R_H^E\right)^2} \right), \nonumber \\
\delta T_E^d &=& - \left(\frac{(d-1) (d-4)}{2}
+ \frac{d_S-d}{\left(d_S-2\right)^2} \frac{\left(d-2\right)^2}{4 (d-1)} \left(3 d^2 -6d -1 +2 (d-1) \left(\psi^{(0)}\left(\frac{2}{d-3}\right) + \gamma \right) \right) \right). \label{tdile}
\eea
From (\ref{deltat}), the absorption cross section comes as
\bea
\sigma &=& A_H^E \left(1 - \delta T_E^d\ \frac{\a}{4 \left(R_H^E\right)^2} \right). \label{seccaodil}
\eea
We have numerically evaluated the $\a$--correction for the cross section: it is always positive, for every relevant value of $d.$

The mass of this black hole in the Einstein scheme is of the form (\ref{massa}), i.e.
\bea
M&=& \left( 1 + \delta M_E^d \ \frac{\a}{4 \left(R_H^E\right)^2}\right) \frac{\left(d-2\right) \Omega_{d-2}}{16 \pi G} \left(R_H^E\right)^{d-3}, \nonumber \\
\delta M_E^d &=& \frac{(d-3)}{4 (d-1) (d_S-2)^2} \left[ 2 (d-1) (d-2)^2 (d-4) - 2 (d-1) (d-2) (d_S-d) \left(\psi^{(0)}\left(\frac{2}{d-3}\right) + \gamma \right) \right. \nonumber \\
&+& \left. (d-2)(d^2 -14 d +17) (d_S-d) + 2(d-1)(d-4)(d_S-d)^2 \right]. \label{massadil}
\eea

When $d\equiv d_S$ the solution studied in this section is equivalent to the previously studied one of Callan--Myers--Perry given by (\ref{fccmp}); in this case, (\ref{tdile}), (\ref{seccaodil}) and (\ref{massadil}) reduce to (\ref{tcmpe}), (\ref{seccaocmp}) and (\ref{mcmpe}), as expected.

\subsection{The doubly charged $d$--dimensional black hole}
\label{giveon}
\indent

In article \cite{Giveon:2009da} one can find black holes in any dimension formed by a fundamental
string compactified on an internal circle with any momentum $n$
and winding $w,$ both at leading order and with leading $\a$
corrections. One starts with the Callan--Myers--Perry solution in the string scheme given in (\ref{fscmp}), (\ref{gscmp}). This metric is lifted to an additional dimension by adding an extra coordinate, taken to be compact (this means to produce a uniform black string). One then performs a boost along this extra direction, with parameter $\alpha_w$, and $T$-dualizes around it (to change string momentum into
winding), obtaining a $(d+1)$--dimensional black string winding around a circle. Finally one boosts one other time along this extra direction, with parameter $\alpha_p$, in order to add back momentum charge. One finally obtains a spherically symmetric black hole in $d$ dimensions with two electrical charges.

The whole process is worked out in detail in \cite{Giveon:2009da}; the final metric, in the string scheme, is of the form (\ref{schwarz}), with $f, g$ given by
\bea
f_S(r)&=&\frac{f_0^T}{\Delta(\alpha_n)\Delta(\alpha_w)}
\left[1+\frac{\a}{2 \left(R_H^S\right)^2} \frac{\mu(r)}{\Delta(\alpha_n)\Delta(\alpha_w)}
-\frac{\a}{2 \left(R_H^S\right)^2} \mu(r) \frac{\sinh^2(\alpha_n)\sinh^2(\alpha_w)}{\Delta(\alpha_n)\Delta(\alpha_w)} \left(\frac{R_H^S}{r}\right)^{2(d-3)}\right. \label{fs} \\
&+&\frac{\a}{2 \left(R_H^S\right)^2} \mu(r)\left(\frac{\sinh^{2}\alpha_n}{\Delta(\alpha_n)}
+\frac{\sinh^{2}\alpha_w}{\Delta(\alpha_w)}\right)+\left. \frac{\a}{4 \left(R_H^S\right)^2} (d-3)^2 f_0^T \left(\frac{R_H^S}{r}\right)^{2(d-2)}
\frac{\sinh^2(\alpha_n)\sinh^2(\alpha_w)}{\Delta(\alpha_n)\Delta(\alpha_w)}\right], \nonumber\\
\Delta\left(x\right)&:=&1+\left(\frac{R_H}{r}\right)^{d-3}\sinh^2x, \label{delta}
\\
g_S(r)&=&f_0^T\left( 1-\frac{\a}{2 \left(R_H^S\right)^2} \epsilon(r)\right) \label{gs}.
\eea
The dilaton in this case is given by

\bea
e^{-2\phi}&=&\sqrt{\Delta(\alpha_n)\Delta(\alpha_w)}
\left[1-2 \frac{\a}{4 \left(R_H^S\right)^2} \varphi(r)-\frac{\a}{4 \left(R_H^S\right)^2} \mu(r) f_0^T \left(\frac{\sinh^2\alpha_n}{\Delta(\alpha_n)}
+\frac{\sinh^2\alpha_w}{\Delta(\alpha_w)}\right) \right. \nonumber\\
&-&\left.\frac{\a}{4 \left(R_H^S\right)^2} \frac{(d-3)^2}{2} f_0^T \left(\frac{R_H^S}{r}\right)^{2(d-2)}
\frac{\sinh^2(\alpha_n)\sinh^2(\alpha_w)}{\Delta(\alpha_n)\Delta(\alpha_w)}\right], \label{conf}
\eea
with $\varphi(r)$ still given by (\ref{fr2}).

In order for the functions $f,g$ to have the form (\ref{fcgc}), one could take a conformal transformation of the metric, changing scheme: $g_{\mu\nu}^I = e^{-2\phi} g_{\mu\nu}^S,$ $e^{-2\phi}$ being given by (\ref{conf}). In this case $f_0$ would have the form (\ref{tangherc}), with $f_0^I(r)=\frac{f_0^T(r)}{\sqrt{\Delta(\alpha_n)\Delta(\alpha_w)}}$ and $c(r)=\frac{1}{\sqrt{\Delta(\alpha_n)\Delta(\alpha_w)}}.$ It would be easy to obtain the functions $f_c, g_c,$ according to (\ref{fcgc}). But with this procedure we would not obtain a metric of the form (\ref{schwarz}), since the form of the $r^2$ factor in front of $d\Omega_{d-2}^2$ would not be preserved. We could solve that by defining a new radial coordinate as $r_I= e^{-\phi (r)} r,$ but that would imply to write the metric in terms of $d\ r_I^2$ instead of $d\ r^2,$ and this way we would loose the assumed form (\ref{fcgc}) for the functions $f,g.$

Instead, it is more convenient to consider the original string scheme metric given in terms of the functions $f_S(r), g_S(r)$ in (\ref{fs}), (\ref{gs}) and carefully look at their near--horizon limit. From (\ref{gscmp}) we see that $g_S(r)\equiv g^{CMP}_S(r).$ From (\ref{delta}) we get $\Delta(x) \underset{r \rightarrow R_H}{\longrightarrow} \cosh^2 x$ and this way
\bea
\frac{f_S(r)}{f_0^T(r)}&\underset{r \rightarrow R_H^S}{\longrightarrow} &
\left[1 +\frac{\a}{2 \left(R_H^S\right)^2} \mu\left(R_H^S\right) \left(\frac{1}{\cosh^2 \alpha_n \cosh^2 \alpha_w} - \tanh^2 \alpha_n \tanh^2 \alpha_w +\tanh^2 \alpha_n + \tanh^2 \alpha_w \right) \right]
\nonumber \\
&\times& \frac{1}{\cosh^2 \alpha_n \cosh^2 \alpha_w } \nonumber
\eea
Simplifying the above expression using $\cosh^2 x - \sinh^2 x =1$ we see that, close to the horizon, we have $f_S(r) \simeq c(R_H^S)^2 f^{CMP}_S(r),$ with $f^{CMP}_S(r)$ being given by (\ref{fscmp}) and $c(R_H^S) = \frac{1}{\cosh \alpha_n \cosh \alpha_w}.$ This means that, re-scaling the time as $d\tilde{t}=c(R_H^S) \, dt$ (in a procedure analogous to the one which we took after (\ref{deltat}), when we mentioned that time and temperature could be always chosen in order to set $c(R_H) \equiv 1$), near the horizon the doubly charged black hole we have been analyzing is written as
$$ ds^2 = -f^{CMP}_S(r)\ d\tilde{t}^2  + g^{CMP}_S(r) \ dr^2 + r^2 d\Omega^2_{d-2},$$
which is exactly the metric of the Callan--Myers--Perry solution written in the string scheme. This way, there is a system of coordinates such that the near horizon geometry of this black hole solution in the string scheme is the same as the Callan--Myers--Perry solution, and therefore so is the black hole temperature, given by (\ref{tcmps}). From (\ref{deltat}) we obtain the absorption cross section:
\be
\sigma = A_H^S \left(1 - \frac{\a}{4 \left(R_H^S\right)^2} \delta T_S^{CMP}\right).\label{seccaogorb}
\ee
with $\delta T_S^{CMP}$ defined in (\ref{tcmps}) and $A_H^S = \left(R_H^S\right)^{d-2} \Omega_{d-2}.$
We have again numerically evaluated the $\a$--correction for the cross section: like in the previous cases, it is always positive, for every relevant value of $d.$

\section{Comparison between the black hole cross section and entropy}
\label{entropia}
\indent

As we have seen, in classical Einstein gravity the low frequency limit of the absorption cross section of minimally coupled massless fields, for any spherically symmetric black hole in arbitrary $d$ dimensions, equals the area of the black hole horizon \cite{dgm96}. In terms of a physical quantity, the Bekenstein--Hawking entropy, this statement may be written as $\left.\sigma\right|_{\a=0} = 4 G \left.S\right|_{\a=0}.$

It is an interesting physical question to figure out if such relation is preserved in the presence of $\a$ corrections, i.e. to verify if the corrections to the cross sections we have been obtaining and to the black hole entropy are the same. The $\a$--corrected entropy can be obtained through Wald's formula
\be
S=-2 \pi \int_H \frac{\partial
\mathcal{L}}{\partial \R^{\mu \nu \rho \sigma}}
\varepsilon^{\mu\nu} \varepsilon^{\rho\sigma} \, \sqrt{h} \,
d\,\Omega_{d-2}, \label{wald0}
\ee
$\mathcal{L}$ being the lagrangian one is considering, which can include $\a$ corrections; $H$ is the black
hole horizon, with area $A_H$ and metric $h_{ij}$ induced by the spacetime metric $g_{\mu\nu}.$ $\varepsilon^{\mu\nu}$ is the binormal to $H.$

For the metric (\ref{schwarz}) we are considering, the nonzero components of $\varepsilon^{\mu\nu}$ are $\varepsilon^{tr}=-\varepsilon^{rt}=-\sqrt{\frac{g}{f}}.$ For the $\a$--corrected lagrangian (in the Einstein scheme) (\ref{eef}) we took, we have
$$8 \pi G\frac{\partial \mathcal{L}}{\partial \R^{\mu \nu \rho
\sigma}}=\frac{1}{4}\left(g_{\mu\rho}g_{\sigma\nu}-g_{\mu\sigma}g_{\rho\nu}\right)+
\mbox{e}^{\frac{4}{d-2} \phi} \frac{\lambda'}{2} \frac{\partial
Y(\R)}{\partial \R^{\mu \nu \rho \sigma}}.$$
This way, taking only nonzero components, one gets from
(\ref{schwarz})
\be
8 \pi G \frac{\partial \mathcal{L}}{\partial
\R^{\mu \nu \rho \sigma}} \varepsilon^{\mu\nu}
\varepsilon^{\rho\sigma} = 4 \times 8 \pi G \frac{\partial
\mathcal{L}}{\partial \R^{trtr}} \varepsilon^{tr} \varepsilon^{tr}
=\left(-\frac{f}{g}+\mbox{e}^{\frac{4}{d-2} \phi} 2 \lambda' \frac{\partial
Y(\R)}{\partial \R^{trtr}} \right)\frac{g}{f}, \label{wald}
\ee
and therefore
\be
S=\frac{1}{4 G} \int_H \left( 1 -2 \lambda' \frac{\partial
Y(\R)}{\partial \R^{trtr}} \right) \, \sqrt{h} \, d\,\Omega_{d-2} =\frac{A_H}{4 G}
-\frac{\lambda'}{2 G} \int_H \frac{\partial
Y(\R)}{\partial \R^{trtr}} \, \sqrt{h} \, d\,\Omega_{d-2}.
\ee

Here one should notice that the $\lambda'=0$ part of the integrand could in principle also contribute to the $\a$--correction to the entropy, because of the $\a$--correction to the metric. But this $\lambda'=0$ part is actually constant, as one can see from (\ref{wald}), no matter what $f, g$ actually are. This way, the $\a$--correction to the entropy depends only on the $\lambda'$--correction term in (\ref{wald}) which, to first order in $\lambda',$ should be computed with the $\lambda=0$ part of the metric. Therefore, the $\a$--correction to the entropy does not depend on the $\lambda'$--corrections to the metric (to first order in $\lambda'$), and we may write
\be
S=\frac{A_H}{4 G}
\left(1+ \lambda \ \delta S \right). \label{entropy}
\ee
Like $\delta M$ in (\ref{massa}) and $\delta T$ in (\ref{temph}), $\delta S$ is a dimensionless factor depending on the specific correction and solution one is considering.

For the case $Y(\R)= \frac{1}{2}\ \R^{\mu\nu\rho\sigma} \R_{\mu\nu\rho\sigma}$ and $\lambda' = \frac{\a}{4},$ corresponding to the particular solutions in the Einstein scheme we have been studying, one has $\frac{\partial Y(\R)}{\partial \R^{trtr}} = \R_{trtr}.$ At order $\a=0,$ $\phi=0,$ $f=g=f_0^T.$ In this case $\R_{trtr} = \frac{1}{2} f''.$ $f''=f_0^{T''}=-\frac{\left(R_H^E\right)^{d-3}}{r^{d-1}} (d-3) (d-2).$ Therefore
\be
S=\frac{1}{4 G} \int_H \left( 1 +\frac{\a}{4 \left(R_H^E\right)^2} (d-3)
(d-2)\right) \, \sqrt{h} \, d\,\Omega_{d-2} =\frac{A_H^E}{4 G}
\left(1+ (d-3) (d-2) \frac{\a}{4 \left(R_H^E\right)^2}\right). \label{pmes}
\ee
This same result was first obtained (by a different process, though) in \cite{cmp89}. From what we have just seen, it is no surprise that the result (\ref{pmes}) is the same for the Callan--Myers--Perry solution (\ref{fccmp}), for the dilatonic solution (\ref{fcdil}) and for any solution for which the near--horizon limit of its classical part is the Tangherlini solution (\ref{tangher}).

The absorption cross section of the doubly charged black hole we considered in section (\ref{giveon}) has been obtained in the string scheme, where its near horizon geometry is that of the Callan--Myers--Perry black hole. We should therefore compute its entropy in the string scheme. Since the entropy, as a physical quantity, is not affected by the change of schemes, it is given by the result in (\ref{pmes}), but with $R_H^E$ replaced by $R_H^S.$ This replacement can be made using (\ref{rhes}), with $\delta T_E$ given by $\delta T_E^{CMP}$ defined in (\ref{tcmpe}) and $\delta T_S$ given by $\delta T_S^{CMP}$ defined in (\ref{tcmps}). The final result is
\be
S=\frac{A_H^S}{4 G}
\left(1+ (d-2) \left(\delta T_S^{CMP} -\frac{(d-2)(d-5)}{2}\right) \frac{\a}{4 \left(R_H^S\right)^2}\right). \label{pmess}
\ee

Comparing the value obtained in (\ref{pmes}) with those in (\ref{seccaocmp}), (\ref{seccaodil}) and the one in (\ref{pmess}) with (\ref{seccaogorb}), we see that in every case we have $\delta S \neq - \delta T.$ This way we conclude that the $\a$ corrections to the absorption cross section and to the entropy do not coincide, for a generic black hole solution.

\section{Covariant and scheme--independent formulae for the black hole entropy and absorption cross section}
\label{ecspv}
\indent

As we previously saw, the $\a$ correction factor $-\delta T$ in (\ref{deltat}) is not invariant under field (namely metric) redefinitions: its value depends on the scheme we take to compute it. Also precisely because of such term giving the $\a$ correction, the expression (\ref{deltat}) is not covariant. Indeed the results we obtained for the cross section, using (\ref{deltat}), are valid only for a particular system of coordinates, namely in which the horizon radius $R_H$ has no $\lambda$ corrections. Since for each of the cases we considered the entropy and the cross section have been obtained using this same system of coordinates, and in the same scheme, it is legitimate to compare their values and to conclude that their $\a$ corrections are not the same, as we did in the previous section. But it would be clearly useful to obtain expressions for both the absorption cross section (\ref{seccao}) and the entropy which are both covariant and invariant under field redefinitions: that could clarify if there exists (or not) a relation between these two quantities; nonetheless, it is certainly more convenient to express them in terms of other quantities which do not depend on systems of coordinates or metric redefinitions. That is also not the case of the horizon area (although this is a covariant quantity).

Suitable quantities for this purpose are the black hole mass and temperature, given for metrics of the form (\ref{schwarz}) respectively by (\ref{massa}) and (\ref{temph}). One can invert these relations in order to obtain in each case the black hole radius $R_H(M), \, R_H(T):$
\bea
R_H(M)&=&\frac{1}{\sqrt{\pi}} \left(\frac{8 G\, M \Gamma\left(\frac{d-1}{2}\right)}{d-2}\right)^{\frac{1}{d-3}} \left[1-\frac{\lambda}{(d-3)} \delta M \right], \label{rhm} \\
R_H(T)&=& \frac{d-3}{4 \pi T} \left(1+ \lambda \delta T \right). \label{rht}
\eea
Replacing $R_H(M)$ (respectively $R_H(T)$) in $A_H=R_H^{d-2} \Omega_{d-2},$ we can obtain the horizon area as a function of the black hole mass (respectively temperature). Replacing these results in (\ref{entropy}), we get
\bea
S(M)&=& 2^{\frac{2d-3}{d-3}} \sqrt{\pi} \left(G \Gamma\left(\frac{d-1}{2}\right)\right)^{\frac{1}{d-3}} \left(\frac{M}{d-2}\right)^{\frac{d-2}{d-3}} \left[1+\lambda\left(\delta S - \frac{d-2}{d-3} \delta M \right) \right], \label{sm} \\
S(T)&=&  \frac{\Omega_{d-2}}{4 G} \left(\frac{d-3}{4 \pi T}\right)^{d-2} \left(1+ \lambda \left(\delta S + (d-2) \delta T \right) \right). \label{st}
\eea

In order to better illustrate the procedure, first we will consider the Callan--Myers--Perry solution. Considering the Einstein scheme values for this solution $\delta M_E^{CMP}, \delta S_E^{CMP}$ respectively from (\ref{mcmpe}), (\ref{pmes}), and the corresponding string scheme values $\delta M_S^{CMP}, \delta S_S^{CMP}$ respectively from (\ref{mcmps}), (\ref{pmess}), it is easy to verify that we have $\delta S_E^{CMP} - \frac{d-2}{d-3} \delta M_E^{CMP} = \delta S_S^{CMP} - \frac{d-2}{d-3} \delta M_S^{CMP}.$ In both cases, replacing that result in (\ref{sm}) we obtain the same expression,
\be
S(M) = 2^{\frac{2d-3}{d-3}} \sqrt{\pi} \left(G \Gamma\left(\frac{d-1}{2}\right)\right)^{\frac{1}{d-3}} \left(\frac{M}{d-2}\right)^{\frac{d-2}{d-3}} \left[1+\a \frac{(d-2)^2}{8} \pi \left(\frac{d-2}{8 G\, M \Gamma\left(\frac{d-1}{2}\right)}\right)^{\frac{2}{d-3}} \right],
\ee
which indeed represents the entropy of the Callan--Myers--Perry black hole as a function of its mass, to first order in $\a$, and is a scheme--independent function.

Still for the same solution, considering the Einstein scheme values $\delta T_E^{CMP}, \delta S_E^{CMP}$ respectively from (\ref{tcmpe}), (\ref{pmes}), and the corresponding string scheme values $\delta T_S^{CMP}, \delta S_S^{CMP}$ respectively from (\ref{tcmps}), (\ref{pmess}), we can also verify that $\delta S_E^{CMP} + (d-2) \delta T_E^{CMP} = \delta S_S^{CMP} + (d-2) \delta T_S^{CMP}.$ In both cases, replacing that result in (\ref{st}) we also obtain the same expression,
\be
S(T) = \frac{\Omega_{d-2}}{4 G} \left(\frac{d-3}{4 \pi T}\right)^{d-2} \left(1 - \a \frac{(d-2)^2 (d-5)}{8} \left(\frac{4 \pi T}{d-3}\right)^2 \right),
\ee
which represents the entropy of the Callan--Myers--Perry black hole as a function of its temperature, to first order in $\a$, and is also a scheme--independent function. Similar expressions $S(M), S(T)$ can be obtained for the dilatonic \cite{Moura:2009it} and doubly charged \cite{Giveon:2009da} black holes, replacing in (\ref{sm}) and (\ref{st}) the corresponding values of the $\a$ corrections.

One can follow exactly the same procedure and replace $R_H(M)$ (respectively $R_H(T)$) in the horizon area in (\ref{deltat}), obtaining what would be the absorption cross section of a spherically symmetric black hole, as a function of its mass (respectively temperature):
\bea
\sigma(M)&=&2^{\frac{4 d-9}{d-3}} \sqrt{\pi} \left(\Gamma \left(\frac{d-1}{2}\right)\right)^{\frac{1}{d-3}} \left(\frac{GM }{d-2}\right)^{\frac{d-2}{d-3}} \left(1-\lambda \left(\delta T + \frac{d-2}{d-3} \delta M \right) \right), \label{seccaom} \\
\sigma(T)&=&\left(\frac{d-3}{4 \pi T}\right)^{d-2} \Omega_{d-2} \left(1+ (d-3) \lambda \delta T \right). \label{seccaot}
\eea
But for the cross section, concerning the Callan--Myers--Perry black hole, if one replaces in (\ref{seccaom}) the Einstein scheme values one obtains a different result than if one replaces the string scheme values: $\delta T_E^{CMP} + \frac{d-2}{d-3} \delta M_E^{CMP} \neq \delta T_S^{CMP} + \frac{d-2}{d-3} \delta M_S^{CMP}.$ Without surprise, and quite obviously, if one analogously replaces in (\ref{seccaot}) the Einstein scheme value one obtains a different result than if one replaces the string scheme value: $\delta T_E^{CMP} \neq \delta T_S^{CMP}.$ This means that, differently from the black hole entropy, the string--corrected absorption cross section cannot be expressed exclusively as a function of the black hole mass (or temperature) in a way which is independent of metric redefinitions, i.e. of the chosen scheme.

From a computational point of view, comparing the $\lambda$ correction terms in (\ref{sm}), (\ref{st}) with those in (\ref{seccaom}), (\ref{seccaot}), that impossibility is easy to understand. Since classically, up to a factor of $4 G,$ both expressions are equal to the horizon area, they represent the same power of $R_H$, which provides the same factor in front of the $\delta T$ or $\delta M$ terms, plus an intrinsic term, which is $\delta S$ (given in (\ref{entropy})) for the entropy, and $\delta \sigma$ for the cross section. The entropy can be expressed in terms of the mass or temperature: the specific values of the intrinsic correction $\delta S$ in the different schemes allow for that. Given that fact, for the same to be possible with the cross section, the intrinsic correction $\delta \sigma$ would have to equal $\delta S$, or at most their difference would have to be a scheme--independent constant. But (\ref{deltat}) tells us that $\delta \sigma = - \delta T$, and one can easily check that $\delta S_E^{CMP} + \delta T_E^{CMP} \neq \delta S_S^{CMP} + \delta T_S^{CMP}.$ Once more, this leads to the impossibility we just mentioned. The reason for this impossibility lies precisely in the string $\a$ corrections and their coefficients: classically, those expressions in terms of mass or temperature are possible. We would have found the same impossibility if we had taken the $\lambda$ corrections corresponding to the other black holes we considered.

Thinking independently of these solutions and considering just the classical terms, without the $\lambda$ corrections, from (\ref{temph}) and (\ref{massa}) we obtain $\left.\sigma\right|_{\a=0} = \left.A_H\right|_{\a=0} = 4 G \, \frac{d-3}{d-2} \frac{\left.M\right|_{\a=0}}{\left.T\right|_{\a=0}};$ including the $\lambda$ corrections, namely from (\ref{deltat}), we see that such classical expression does not generalize. Nonetheless, the presence of the $1/T$ factor seems to indicate the right dependence on the temperature that the cross section should have, in order to naturally absorb the $\lambda$ term in (\ref{deltat}) (this had already been suggested in appendix \ref{fceqgc}). Replacing in (\ref{deltat}) one of the factors of $R_H$ by $R_H(T)$ given in (\ref{rht}), we obtain to first order in $\lambda$ $\sigma=R_H^{d-2} \Omega_{d-2} \left(1- \lambda \delta T\right) = \frac{d-3}{4 \pi T} R_H^{d-3} \Omega_{d-2},$ i.e.
\be
\sigma = \frac{d-3}{4 \pi T} \Omega_{d-2}^{\frac{1}{d-2}} A_H^{\frac{d-3}{d-2}}. \label{seccaoto}
\ee
Since it was obtained only from (\ref{deltat}) and (\ref{rht}), (\ref{seccaoto}) represents an expression for the absorption cross section which is covariant and valid on every scheme, for generic spherically symmetric $d-$dimensional black holes, to first order in $\lambda.$

Several interesting questions can be raised. We can only guarantee that (\ref{seccaoto}) is valid to first order in  $\lambda,$ because that is the order we worked with in our derivation. Interestingly there is no explicit dependence on $\lambda$ (or $\a$) in this expression. Could it still be valid to higher (maybe arbitrary) orders in $\a?$ Also, could an expression like (\ref{seccaoto}) be valid for non--spherically symmetric black holes?

The $1/T$ dependence of $\sigma$ in (\ref{seccaoto}) raises the issue of its validity in the $T \rightarrow 0$ limit, i.e. for extremal black holes (we recall that in our derivation we assumed we were dealing with nonextremal black holes). That could be combined with the $A_H \rightarrow 0$ limit: the case of small black holes, where the horizon area vanishes classically and is of order $\lambda.$ These questions are to be addressed in future works.

\section{Discussion and future directions}
\indent

In this article, we have obtained a general formula for the low frequency absorption cross section for spherically symmetric $d$--dimensional black holes with leading $\a$ corrections in string theory, which we applied to known black hole solutions. First we obtained it in a form (\ref{deltat}) that can be applied in a specific scheme and coordinate system, but later we wrote it in a form (\ref{seccaoto}) which is covariant and scheme--independent, given in terms of the black hole area and temperature.

A remarkable fact about our $\a$--corrected cross section, either in the forms (\ref{deltat}) or (\ref{seccaoto}), is that it depends exclusively on information computed at the black horizon. Indeed, as we have seen, only the $\lambda=0$ contribution from the intermediate region affects the matching (and the final result). At asymptotic infinity the analysis is exactly the same as without $\a$ corrections, and close to the horizon, the potential $V[f(r),g(r)]$ given by (\ref{potential0}) also vanishes and the only effect of the $\a$ corrections comes from the approximation (\ref{fghor}).

This suggests some kind of universality: maybe the low frequency limit of the cross section is the same not only for minimally coupled massless scalar fields, but also for other types of fields. In particular, our result for the cross section finally does not depend on the effective potential and on the $\a$ corrections that it may contain. This fact, together with the absence of explicit $\a$ corrections in (\ref{seccaoto}), allows us to go even further and propose that a result like (\ref{seccaoto}) could be valid to higher orders in $\a.$ All these claims should be checked in future works. A first step would be the generalization of the cross section formula (\ref{seccao}) to include next to leading order $\a$ corrections.

Another general question we have addressed in this work is the validity of the relation $\sigma= 4 G S$ in the presence of $\a$ corrections. Indeed, from our discussion of section \ref{entropia}, we showed that the entropy, to first order in $\a,$ depended exclusively on the classical $\a=0$ metric, while from (\ref{seccao}) the absorption cross section depends explicitly on the $\a$ corrections to the metric. The examples we have analyzed confirm such discrepancy, which can be understood with the analysis of section \ref{ecspv}. Indeed, as we saw the $\a$--corrected entropy can be expressed in terms of the mass or temperature, something which is not possible with the $\a$--corrected absorption cross section. This fact clearly shows that the entropy and the (low frequency) cross section are two distinct quantities; the fact that, classically, they are related (up to a factor $4 G$) is just a coincidence, at least in string theory.

But there are examples in the literature where the agreement $\sigma= 4 G S$ exists up to higher orders in $\a.$ In article \cite{Cornalba:2006hc} such agreement was found, to all orders in $\a,$ for fundamental strings in the (small) black hole phase (BPS states of heterotic strings compactified on $S^1 \times T^5$). In this article we did not deal with fundamental strings or small black holes, but this gives us a hint that, in some special cases, the agreement may exist.

A more recent example is in article \cite{Kuperstein:2010ka}, where the authors analyzed 1/4 BPS black holes in ${\cal N}=4$ string theory both in $d=4$ and $d=5,$ having in both cases obtained the agreement $\sigma= 4 G S$ just to first order in $\a.$ The examples we have analyzed here are not supersymmetric and are in generic $d$ spacetime dimensions. But the agreement found in \cite{Kuperstein:2010ka} allows us to ask a few questions, which for now remain open: does that agreement hold for generic $d$ dimensions? Does it only hold for supersymmetric black holes? What could be the minimal amount of supersymmetry for it to eventually hold? We cannot provide answers to such questions because, as we have mentioned, our results only apply to non--extremal black holes. In a forthcoming work we will extend the results of this article to extremal (and, in particular, to supersymmetric) black holes.

From what we have seen, the possibility of the two quantities $\sigma, \, S$ having the same correction would require some relation between the classical $\a=0$ metric and its $\a$ corrections. That should also be the object of further study.

\section*{Acknowledgments}
The author wishes to acknowledge useful discussions with Dan Gorbonos, Miguel Paulos, Sameer Murthy and Ricardo Schiappa. This work has been supported by FEDER funds through \emph{Programa Operacional Fatores de Competitividade -- COMPETE} and by Funda\c c\~ao para a Ci\^encia e a Tecnologia through projects Est-C/MAT/UI0013/2011, CERN/FP/116377/2010 and CERN/FP/123609/2011.

\appendix

\section{The $\a$ corrections to the temperature}
\label{fceqgc}
\indent

In order to compute the temperature $T$ of a black hole given by a metric of the form (\ref{schwarz}), one first
Wick--rotates to Euclidean time $t = i \tau$; the resulting manifold has no conical singularities as long
as $\tau$ is a periodic variable, with a period $\beta= \frac{1}{T}.$ The precise smoothness condition is $2 \pi =
\lim_{r \to R_H}{\frac{\beta}{g^{-\frac{1}{2}} \left(r\right)} \frac{d f^{\frac{1}{2}}\left(r\right)}{d r}},$ from which one gets
$$T= \lim_{r \to R_H}{\frac{\sqrt{g}}{2\pi} \frac{d\,\sqrt{f}}{d\,r}}.$$
In the case $f, g$ are given by (\ref{fcgc}), the temperature comes as
\be
T=\frac{f_0'(R_H)}{4 \pi} \left(1+ \lambda \ \frac{f_c(R_H) + g_c(R_H)}{2} \right). \label{temp}
\ee

The $\a$ correction to the temperature is the same obtained to the absorption cross section in (\ref{seccao}), but with opposite sign: when one of these quantities increases, the other one decreases by the same (relative) magnitude. This means the product $\sigma T$ does not get $\a$ corrections to first order.

\section{An example: the potential for tensor--type gravitational perturbations with leading $\a$ corrections}
\indent

As we mentioned in the main text, without $\a$ corrections the equation describing tensor--type gravitational perturbations of a spherically symmetric metric in $d$
dimensions like (\ref{schwarz}) is the same as the field equation for $\mathcal{H}$ (the same being true for the potential $V_{\textsf{cl}} [f(r),g(r)]$, given by
(\ref{potentialcl})). That does not necessarily need to be the case in the presence of $\a$ corrections. Just as an example of a higher order potential, here we show the
potential for tensorial perturbations of a metric like (\ref{schwarz}), but with leading $\a$ corrections quadratic in the Riemann tensor, in the context of heterotic string
theory, i.e. a solution coming from the action (\ref{eef}), with $\lambda' Y(\R)= \frac{\a}{8}\ \R^{\mu\nu\rho\sigma} \R_{\mu\nu\rho\sigma}.$

In a different work \cite{Moura:2012fq} we showed that in this case the perturbation variable obeys an equation like (\ref{dottigen}), with \footnote{It is easy to see that,
for a solution like (\ref{fcgc}), one has $F_{\textsf{T}} = F_{\textsf{cl}},$ as previously mentioned.}
\bea
F_{\textsf{T}} &=& \sqrt{fg}\left(1 +\frac{\a}{4} \frac{f'-g'}{r}\right), \nonumber \\
P_{\textsf{T}} &=& - f \left[ (d-2) \frac{g}{r} + \frac{1}{2} \left(f'+g'\right) + \frac{\a}{4 r^2}  \left(4 (d-4) \frac{g(1-g)}{r}+r g'\left(f'-g'\right)-4 g g' +2 (d-2) g
f' \right) \right], \nonumber \\
Q_{\textsf{T}} &=& \frac{\ell \left( \ell + d - 3 \right)}{r^2} f + \frac{(g-f)f'}{r} +
\frac{\a}{2 r^2} \left[ \frac{\ell \left( \ell + d - 3 \right)}{r} f \left( 2 \frac{1-g}{r} + f' \right) + (g-f) f'^2 \right].
\label{fpqt}
\eea
From (\ref{fpqt}) and (\ref{potentialcl}) we see that the corresponding potential is given by
\bea
V_{\textsf{T}} [f(r),g(r)] &=& V_{\textsf{cl}} [f(r),g(r)] +
\frac{\a}{32 r^4 f g} \left[32 \ell (\ell +d-3) f^2 (1-g) g +16 \ell (d+\ell -3) f^2 g f' r \right. \nonumber \\
&+& 3 r^3 g^2 f'^2 \left(f'-g'\right)-2 r^3 f g f' \left(f'-g'\right) g' -4 r^3 f^2 g f' \left(f''-g''\right)-2 r^3 f^2 g g' \left(f''-g''\right) \nonumber \\
&+& 2 r^3 f g^2 \left(-3 f' f''+2 g' f''+f' g''\right) -4 r^3 f^2 g^2 \left(f^{(3)}-g^{(3)}\right) +18 r^2 f g^2 f'^2 -12 r^2 f^2 g f'^2 \nonumber \\
&-& 10 r^2 f^2 g g'^2 -2 r^2 f g^2 f' g' +2 r^2 (4 d-13) f^2 g f' g' +8 r^2 f^2 g^2 f'' +8 (d-5) r^2 f^2 g^2 g''
 \nonumber \\
&+& 4 r(d-4)^2 f^2 g^2 (f' + g') +8r f^2 g^2(g'-f') +8 (d-4) r f^2 g (f' + g'-4g g') \nonumber \\
&+& \left. 16 (d-5) (d-4) f^2 g^2 (1-g) -r^3 f^2 f'^2 \left(f'-g'\right)\right],
\label{potential}
\eea

Close to the horizon $f, g$ are given by (\ref{fghor}),  and the potential $V_{\textsf{T}} [f(r),g(r)]$ comes as
\bea
V_{\textsf{T}} (r) &\simeq& \frac{r-R_H}{2 R_H} \Big[(d-2) f_0'^2(R_H) + \frac{\a}{4 R_H^2} f_0'(R_H) \left[ 4(d-4) R_H f_0'^2(R_H) \right. \nonumber \\
&+& \left(8(d-4) +(3d-10) f_c(R_H) + (d+2)g_c(R_H) \right. \nonumber \\
&+& \left. \left. R_H \left(f_c'(R_H)-g_c'(R_H) \right) \right) f_0'(R_H) + R_H \left(f_c(R_H)-g_c(R_H)\right) f_0''(R_H) \right]\Big]
+ {\mathcal{O}} \left( \left( r-R_H \right)^2 \right). \label{vthor}
\eea
This means at the precise location of the horizon, $V_{\textsf{T}} [f(r),g(r)]$ vanishes; in the nearby region it may be neglected.

In this article we consider asymptotically flat black holes which, at infinity, behave like flat Minkowski spacetime. Here we make the same assumptions as in section
\ref{sai}, namely that  at asymptotic infinity, in the metric (\ref{schwarz}), functions $f(r), g(r)$ tend to the constant value 1 in the limit of very large $r,$ and their
derivatives tend to 0 in the same limit. From (\ref{potential}) we see that, asymptotically, $V_{\textsf{T}} (r)$ behaves  at most as $1/r^2,$ and therefore it vanishes in
the limit $r \rightarrow \infty.$ The leading $\a$ correction behaves as $1/r^4$ and it also vanishes in this limit.

We conclude that the potential $V_{\textsf{T}} [f(r),g(r)]$ given by (\ref{potential}), which is an example of a potential including $\a$ corrections, satisfies all the assumptions we made in sections \ref{sch} and \ref{sai}.


\bibliographystyle{plain}

\end{document}